\documentclass{aa}  
\usepackage{graphicx}
\usepackage{txfonts}
\usepackage{hyperref}
\usepackage{lscape}
\usepackage{threeparttable}
\usepackage{soul}

\defcitealias{2017MNRAS.469..800M}{Paper\,I}
\defcitealias{2018ApJ...853...86B}{Paper\,II}
\defcitealias{2018ApJ...854...45L}{Paper\,III}
\defcitealias{2021MNRAS.505.3549S}{Paper\,IV}
\defcitealias{2022ApJ...930...24G}{Paper\,V}
\defcitealias{2024A&A...688A.180S}{Paper\,VI}

\begin{document} 

\title{The HST Large Programme on $\omega$\,Centauri - VII.\\ The white dwarf cooling sequence}

%\subtitle{}

\author{M.\,Scalco\inst{1,2}\fnmsep\thanks{\email{michele.scalco@inaf.it}}
\and
M.\,Salaris\inst{3,4}
\and
L.\,Bedin\inst{2}
\and
M.\,Griggio\inst{1,2,5}
\and
A.\,Bellini\inst{5}
\and
M.\,Libralato\inst{2}
\and
D.\,Nardiello\inst{6,2}
\and
E.\,Vesperini\inst{7}
\and 
J.\,Anderson\inst{5}
\and 
P.\,Bergeron\inst{8}
\and
A.\,Burgasser\inst{9}
\and
D.\,Apai\inst{10,11}
}

\institute{Dipartimento di Fisica e Scienze della Terra, Università di Ferrara, Via Giuseppe Saragat 1, Ferrara I-44122, Italy
\and
Istituto Nazionale di Astrofisica, Osservatorio Astronomico di Padova, Vicolo dell’Osservatorio 5, Padova I-35122, Italy
\and
Astrophysics Research Institute, Liverpool John Moores University, 146 Brownlow Hill, Liverpool L3 5RF, UK
\and
Istituto Nazionale di Astrofisica, Osservatorio Astronomico d'Abruzzo, Via Mentore Maggini, Teramo I-64100, Italy
\and
Space Telescope Science Institute, 3700 San Martin Drive, Baltimore, MD 21218, USA
\and
Dipartimento di Fisica e Astronomia "Galileo Galilei", Universit{\`a} di Padova, Vicolo dell'Osservatorio 3, Padova I-35122, Italy
\and
Department of Astronomy, Indiana University, Swain West, 727 E. 3rd Street, Bloomington, IN 47405, USA
\and
D{\'e}partement de Physique, Universit{\'e} de Montr{\'e}al, C.P. 6128, Succ. Centre-Ville, Montr{\'e}al, Quebec H3C 3J7, Canada
\and
Department of Astronomy \& Astrophysics, University of California, San Diego, La Jolla, California 92093, USA
\and
Department of Astronomy and Steward Observatory, The University of Arizona, 933 N. Cherry Avenue, Tucson, AZ 85721, USA
\and
Lunar and Planetary Laboratory, The University of Arizona, 1629 E. University Blvd., Tucson, AZ 85721, USA
}

\date{XXX,YYY,ZZZ}
 
\abstract
{We present a study of the white dwarf (WD) cooling sequence (CS) in the globular cluster (GC) Omega\,Centauri (or NGC\,5139, hereafter $\omega$\,Cen), the primary goal of a dedicated Hubble Space Telescope (HST) programme. Our analysis has revealed that the peak at the termination of the WD CS is located at $m_{\rm F606W}=30.1\pm0.2$ (equivalent to $V$$\sim$31). The brighter part of $\omega$\,Centauri's WD CS is consistent with the presence of massive He-core WDs, in agreement with previous HST analyses with ultraviolet and blue filters. Comparative analyses of the WD luminosity function (LF) with theoretical counterparts have shown that a single-age population for the cluster is compatible with the data. However, an analysis of just the WD LF cannot entirely exclude the possibility of an age range, due to uncertainties in the present-day WD mass function, with a star formation history potentially spanning up to 5 billion years, predominantly comprising stars about 13~Gyr old, and with just a minority potentially as young as 8~Gyr. This underscores the need for global spectroscopic and photometric investigations that include simultaneously the WD populations together with the previous evolutionary phases to fully understand the cluster's diverse chemical compositions and ages.}


\titlerunning{The HST Large Programme on $\omega$\,Centauri - VII.}
\authorrunning{M.\,Scalco et al.}
\maketitle

\section{Introduction}\label{Section1}
The white dwarf cooling sequence of globular clusters lies in one of the faintest and least-explored regions of the colour-magnitude diagram (CMD). Deep imaging with HST has reached the peak of the WD number distribution at the faint end of the CS in four GCs: NGC\,6397 \citep{2008AJ....135.2114A}, M\,4 \citep{2009ApJ...697..965B}, 47\,Tucanae \citep{2012AJ....143...11K}, and NGC\,6752 \citep{2019MNRAS.488.3857B,2023MNRAS.518.3722B}. Each one of these GCs hosts multiple stellar populations (mPOPs) characterized by small mean spreads in initial helium abundances \citep{2018MNRAS.481.5098M}, and their WD CSs align with predictions for single-population (WD population) systems \citep{2013ApJ...778..104R,2016MNRAS.456.3729C}.

Omega Centauri is a moderately low-reddening GC which is relatively close ($\sim$5 kpc) to the Sun, making it an ideal target for an efficient study of its faint WD population. It is also one of the most extreme cases of a GC hosting mPOPs; in fact, spectroscopy has disclosed that its stars display a range of [Fe/H], and a very complex pattern of light-element anti-correlations at each [Fe/H] \citep{2012ApJ...746...14M}, while optical CMDs have revealed two main groups of stars with large differences in their initial helium content\citep[Y$\sim$0.40 for the He-rich component;][]{2012AJ....144....5K}, as evidenced by its split main sequence (MS) in optical filters \citep{2004ApJ...605L.125B}. The large range of initial chemical abundances also manifests itself through UV-based CMDs, that  display a main sequence with at least 15 sub-populations \citep{2017ApJ...844..164B}.  

Regarding the WD population, the upper part of the CS in $\omega$\,Cen exhibits a bifurcation into two sequences, as reported by \citet{2013ApJ...769L..32B}: a blue CS populated by standard CO core WDs, and a red CS populated by low-mass WDs with both CO and (mainly) He cores (the presence of a sizable population of He-core WDs was already disclosed by \citealt{monelli} and \citealt{cc07}).

The prevailing hypothesis suggests that the blue WD CS is populated by the final stages of evolution of the He-normal stars of $\omega$\,Cen, while the red WD CS is populated by the final stages of evolution of the He-rich stars. Observing the complete WD CS of this cluster will allow us to investigate the effect of a chemically complex system like $\omega$\,Cen on the termination of the WD CS (that is an age indicator used in stellar population studies), and set independent constraints on the cluster age spread.

This study represents the seventh paper in a series dedicated to the exploitation of an HST multicycle large programme centred on $\omega$\,Cen. The previous publications focused on the analysis of the parallel fields collected as a part of the programme:

\begin{itemize}
    \item \citet[hereafter \citetalias{2017MNRAS.469..800M}]{2017MNRAS.469..800M} focused on the analysis of the mPOPs of $\omega$\,Cen at very faint magnitudes.
    \item \citet[hereafter \citetalias{2018ApJ...853...86B}]{2018ApJ...853...86B} analysed the internal kinematics of these mPOPs.
    \item \citet[hereafter \citetalias{2018ApJ...854...45L}]{2018ApJ...854...45L} presented the absolute proper motion estimate for $\omega$\,Cen.
    \item \citet[hereafter \citetalias{2021MNRAS.505.3549S}]{2021MNRAS.505.3549S} released the astro-photometric catalogue for two parallel fields of the programme.
    \item \citet[hereafter \citetalias{2022ApJ...930...24G}]{2022ApJ...930...24G} presented a set of stellar models designed to investigate low-mass stars and brown dwarfs in $\omega$\,Cen.
    \item Finally, \citet[hereafter \citetalias{2024A&A...688A.180S}]{2024A&A...688A.180S} presented a comprehensive analysis of the radial distribution of the mPOPs of $\omega$\,Cen across an extensive part of the cluster.
\end{itemize}

The focus of this study is the WD CS of the cluster and its LF in the primary deep field of this HST large programme. 

The paper is organized as follows: Section\,\ref{Section2} outlines the observations, while section\,\ref{Section3} gives details about the data reduction. Section\,\ref{Section4} offers a concise overview of the artificial star tests (ASTs) we have performed. The selection criteria employed to construct the cluster CMD are detailed in Section\,\ref{Section5}. The decontamination of the cluster sample using proper motions (PMs) is discussed in Section\,\ref{Section6}. Section\,\ref{Section7} presents the empirical WD LF derived from the data, and Section\,\ref{Section8} discusses the theoretical interpretation of the WD LF. Finally, Section\,\ref{Section9} provides a brief summary of the results.

%__________________________________________________________________
% Memo4me: /D2/wCenWDCS2020/fromSkalko/
\begin{figure*}[t]
\centerline{
\includegraphics[height=57mm]{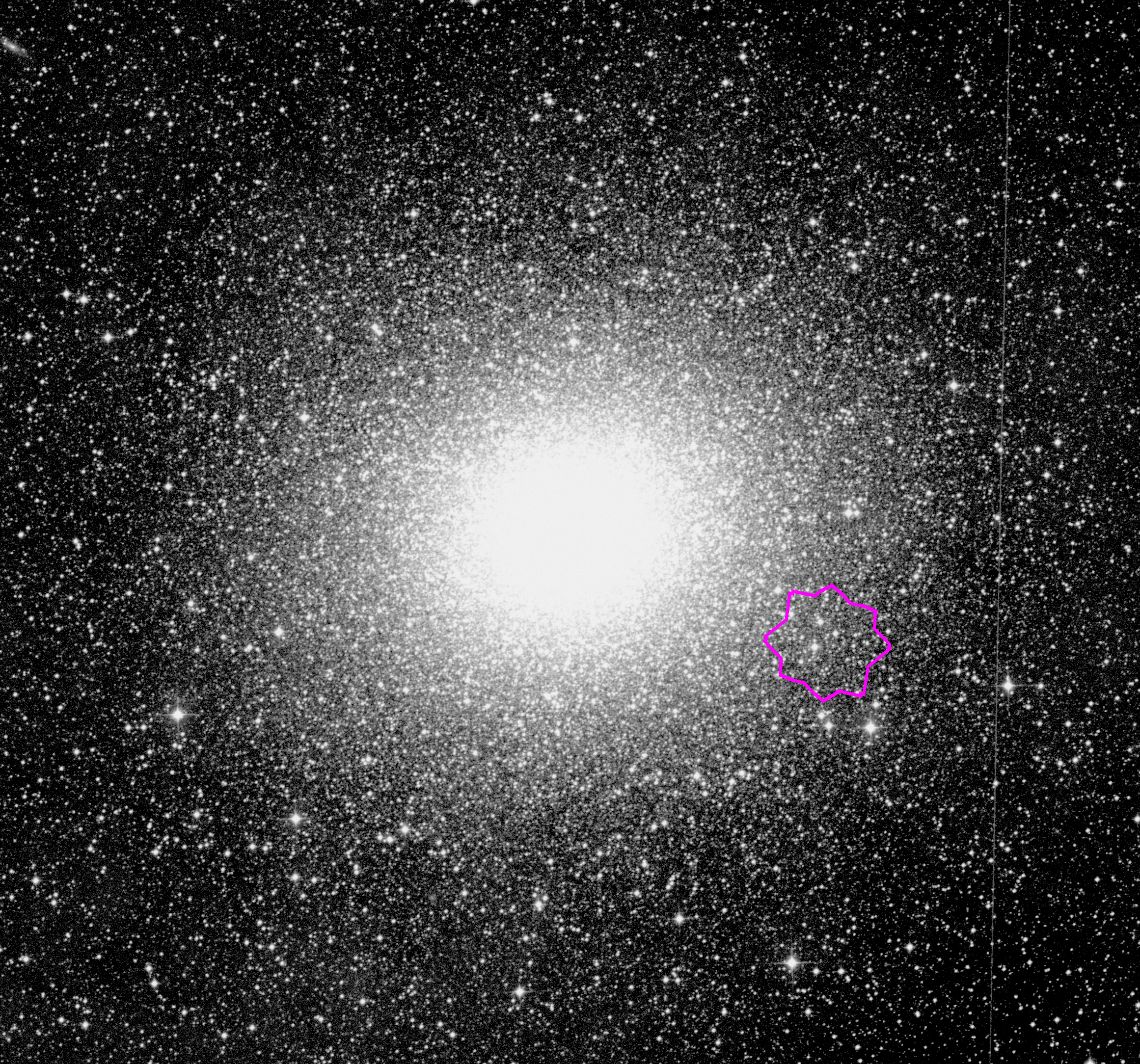}
\includegraphics[height=57mm]{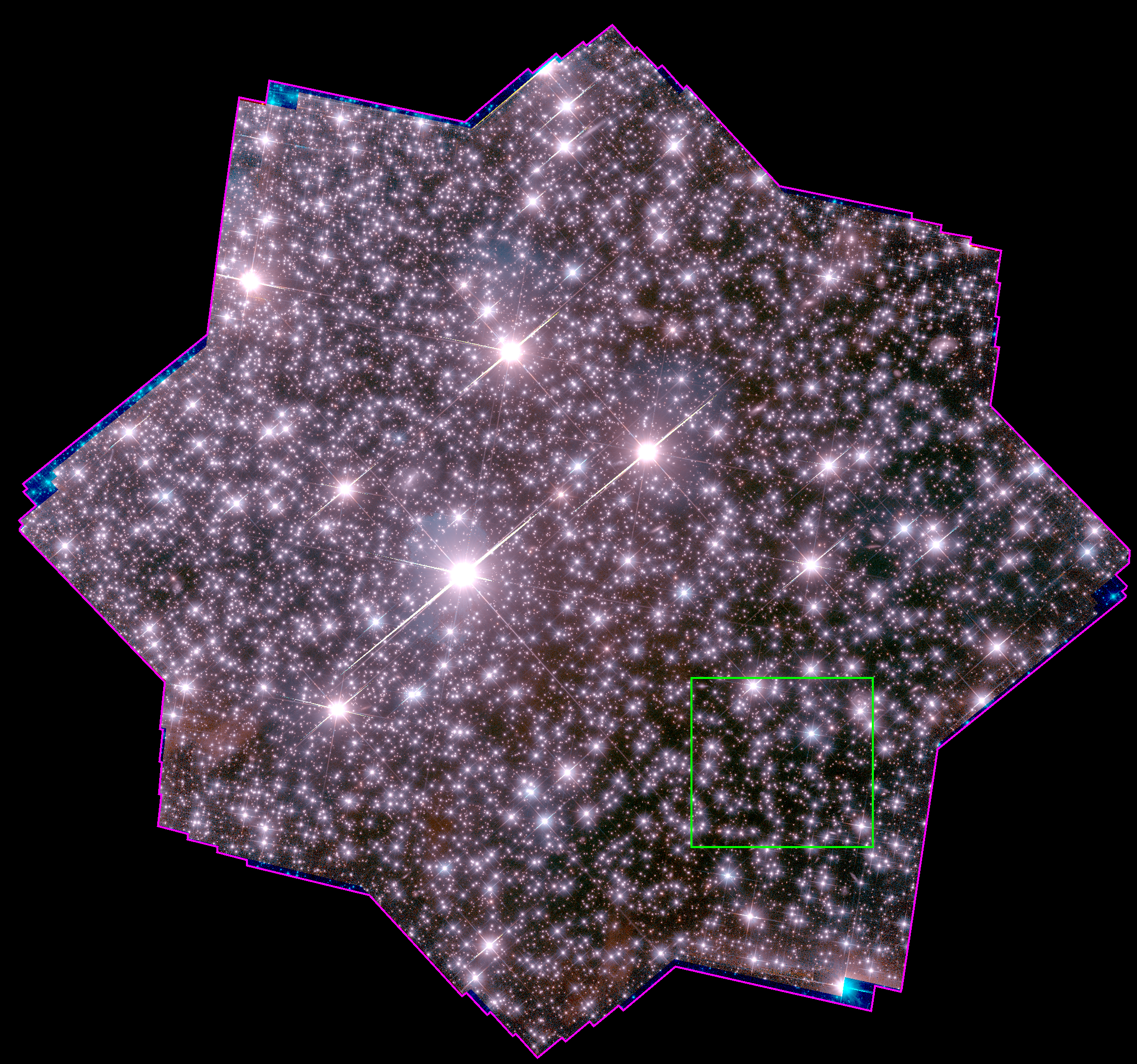}
\includegraphics[height=57mm]{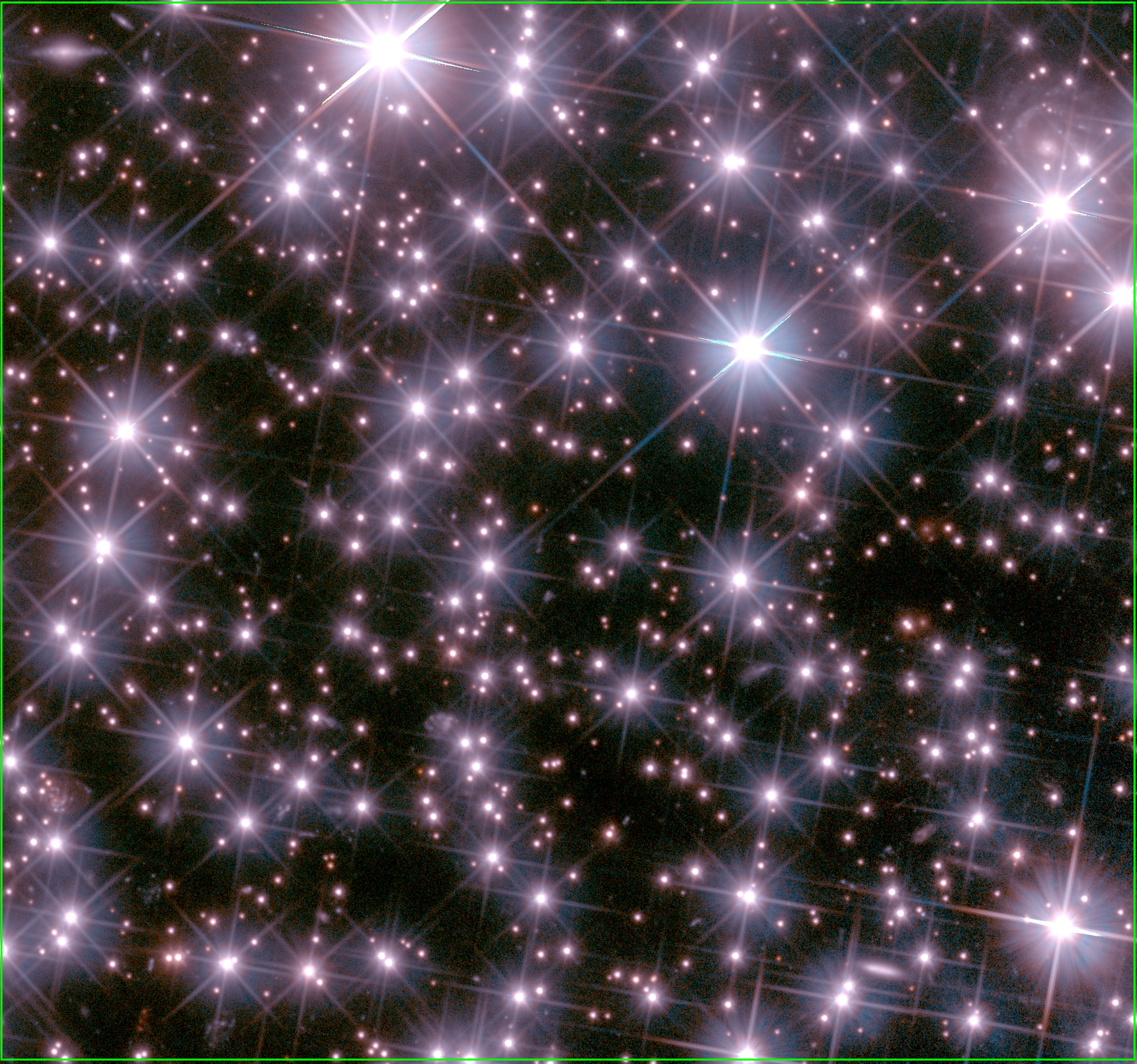}
}
\caption{
%%%
\textit{(Left:)} A $1^\circ\times1^\circ$ infrared image from the Digital\,Sky\,Survey\,2 centered on $\omega\,Cen$. The image is aligned with North up and East toward the left.
The region indicated in magenta is the field of view covered by our HST deep field from programmes GO-14118 and GO-14662. 
\textit{(Middle:)} Three-chromatic stacked image of the entire ACS/WFC primary field ($\sim5^\prime\times5^\prime$).
\textit{(Right:)} A zoom-in on a dark sub-region (of $\sim1^\prime\times1^\prime$) 
of the ACS/WFC primary field (indicated by the green box in the center panel), at a scale where individual pixels are visible. 
}
%%%
\label{fig:FoV}
\end{figure*}
%__________________________________________________________________
%

\section{Observations}\label{Section2}
This study is based on images collected with the Wide-Field Channel (WFC) of the Advanced Camera for Surveys (ACS) onboard HST under the multicycle large programme \textit{The end of the White Dwarf Cooling Sequences of Omega Centauri} (programme IDs: GO-14118 and GO-14662, PI: Bedin). Data were collected over a total of  of 132 HST orbits across six epochs, spanning a temporal baseline of $\sim$3~years (from 2015 to 2018).  In each epoch, we took 24 deep exposures (of $\sim$1100-1200\,s), in the F606W filter and 20 in the F814W filter, resulting in a total of 216 exposures taken with the F606W filter and 180 exposures with the F814W filter.  The data set also includes 12 F606W and 10 F814W short exposures (of $\sim$40-50\,s). A finding chart of the studied field at two scales, and its position relative to the cluster centre is shown in Fig.\,\ref{fig:FoV}. 

\section{Data reduction}\label{Section3}
The data reduction closely follows the procedure outlined in \citet{2019MNRAS.488.3857B,2023MNRAS.518.3722B}, and in other papers of this series (see \citetalias{2018ApJ...853...86B,2021MNRAS.505.3549S,2024A&A...688A.180S}). For a detailed description of the reduction process, we refer readers to those papers.

We initially conducted a \textit{first-pass} analysis, wherein fluxes and positions for relatively bright, unsaturated stars were extracted from each image using the \texttt{FORTRAN} routine \texttt{hst1pass} \citep[see][]{2022wfc..rept....5A}. Each image was analyzed independently to generate a tailored effective Point Spread Function (ePSF), allowing for adjustments to accommodate spatial and temporal variations relative to the library ePSFs provided by \citet{2006hstc.conf...11A}. The tailoring of ePSFs was performed using the method introduced by \citet{2017MNRAS.470..948A} for the ultraviolet and visible (UVIS) channel of the Wide Field Camera 3 (WFC3), later extended to ACS/WFC by \citet{2018acs..rept....8B}. Both positions and fluxes were corrected for the geometric distortion of the detector following the methods outlined by \citet{2006hstc.conf...11A}. We then created a common, distortion-free reference frame, based on cluster members, to which all individual images were linked by using a six-parameter linear transformation.

With the tailored ePSF and transformations obtained during the \textit{first-pass} analysis, we conducted a \textit{second-pass} analysis employing the \texttt{FORTRAN} code \texttt{KS2} introduced by \citet{2016ApJS..222...11S,2017ApJ...842....6B} (also see \citealt{2008AJ....135.2055A}; \citetalias{2018ApJ...853...86B,2018ApJ...854...45L,2021MNRAS.505.3549S}; \citealt{2018MNRAS.481.3382N} for a comprehensive description of the software). The \texttt{KS2} code iteratively identifies, models, and subtracts point sources from the image, initially targeting the brightest sources and progressively addressing fainter sources within the subtraction residuals. This iterative process determines stars' positions and fluxes, alongside crucial diagnostic parameters such as the local sky noise (rmsSKY) and the RADXS parameter \citep[introduced in][]{2009ApJ...697..965B}, which assesses the resemblance of the source flux distribution to that of the ePSF. For a detailed description of these parameters, we direct readers to \citetalias{2018ApJ...853...86B,2021MNRAS.505.3549S}, and \citealt{2018MNRAS.481.3382N}.

Photometry was calibrated to the ACS/WFC Vega-mag system using the procedure outlined in \citet{2005MNRAS.357.1038B}, employing encircled energy and zero points available from STScI\footnote{\url{www.stsci.edu/hst/acs/analysis/zeropoints}}.

In panel\,(a) of Fig.\,\ref{fig:test}, we present a preliminary $m_{\rm F606W}$ versus $m_{\rm F606W}-m_{\rm F814W}$ colour-magnitude diagram (CMD) for sources satisfying |RADXS|<0.1 in both F606W and F814W filters. This CMD is used exclusively to define the fiducial line of the WD CS (see Section\,\ref{Section4}). For the subsequent analysis, a CMD obtained with different selection criteria is employed.

\section{Artificial stars}\label{Section4} 
\begin{figure}
\centering
\includegraphics[width=\columnwidth]{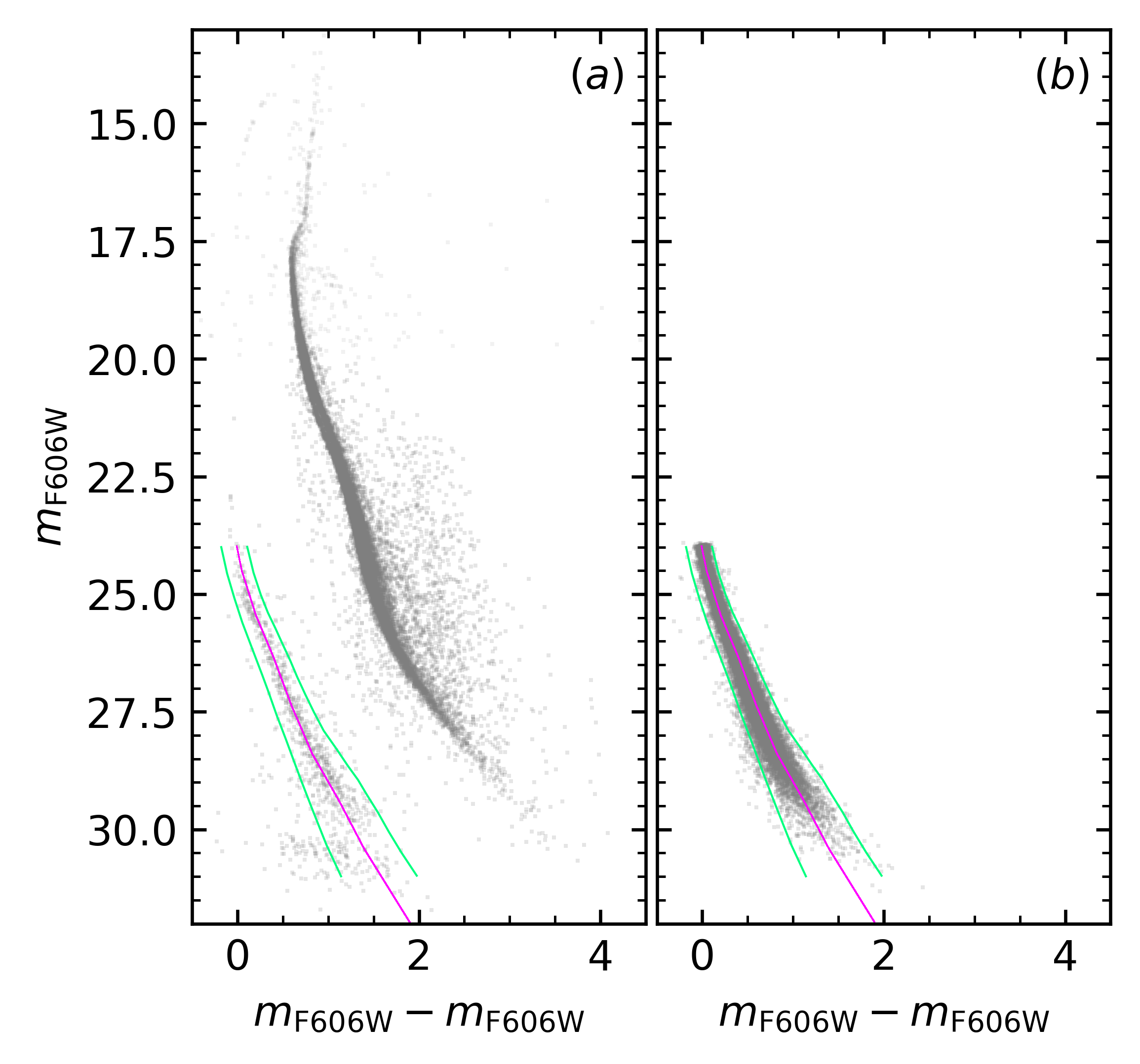}
\caption{(a) Preliminary $m_{\rm F606W}$ versus $m_{\rm F606W}-m_{\rm F814W}$ CMD of sources within the studied field. Only those having |RADXS|<0.1 in both F606W and F814W filters are shown. The magenta line denotes the fiducial line for the WD CS, while the green lines bound the region in the CMD used for counting the WDs. (b) On the same scale, the CMD for ASs is presented. The magenta and green lines from panel\,(a) are retained. The green lines define an area generous enough to encompass both the majority of observed real WDs in panel\,(a) and the ASs introduced along the WD fiducial line, which were recovered despite exhibiting substantial photometric errors.}
\label{fig:test}
\end{figure}

The ASTs were conducted following the guidelines outlined by \citet{2019MNRAS.488.3857B,2023MNRAS.518.3722B}. In summary, a fiducial line was established by hand along the bulk of the observed WDs in the preliminary CMD presented in panel\,(a) of Fig.\,\ref{fig:test}, extending down to where they appeared to stop, and then extrapolated to fainter magnitudes. This fiducial line is shown in magenta in panel\,(a) of Fig.\,\ref{fig:test}.

Using \texttt{KS2}, $10^5$ ASs were introduced along this fiducial line, distributed uniformly with a $m_{\rm F606W}$ magnitude between 24 and 32, and a homogeneous spatial distribution across the field of view. The methodology outlined in \citep[][Section 2.3]{2009ApJ...697..965B} was followed to correct for input-output systematic errors in both real and artificial magnitudes.

To determine if an inserted star was successfully recovered, criteria were set such that if an AS was not detected within 0.753 magnitudes ($\sim -$2.5log 2) in both filters and within 1 pixel from the inserted position in both x and y detector coordinates, it was considered unrecovered. Panel\,(b) of Fig.\,\ref{fig:test} shows the inserted artificial sources (in magenta) and those that were successfully recovered (grey dots).

The combined information from the panels of Fig.\,\ref{fig:test} was used to bound the region used for counting the WDs of $\omega$\,Cen. This region was defined by hand-drawn green lines, striking a balance between encompassing observed WDs with significant photometric scatter and excluding the majority of field objects.

\section{Color-magnitude diagram and selections}\label{Section5}
\begin{figure*}
\centering
\includegraphics[width=\textwidth]{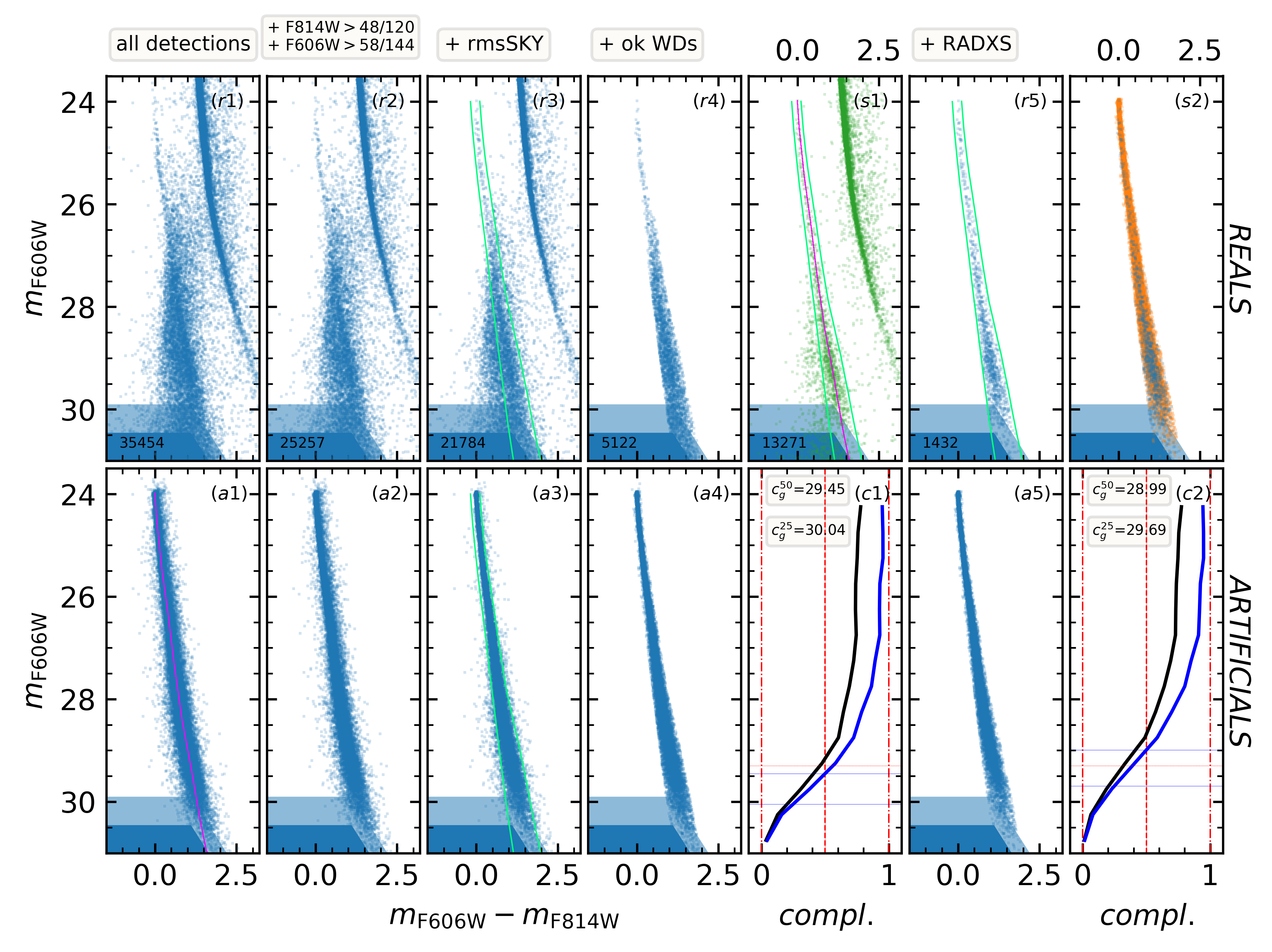}
\caption{Progression of cumulative selections employed to derive a sample of well-measured WDs along $\omega$\,Cen's WD CS. The sequence progresses from left to right, with top panels representing real stars and bottom panels representing ASs. The number of real stars after each selection is reported in the bottom left of the panels. In panel (a1), a magenta line indicates where ASs were introduced. Panels (c) display the resulting completeness, depicted as (c1) without and (c2) with the inclusion of the selection on RADXS, a highly effective parameter for identifying well-measured point sources. Panel (s1) illustrates the impact of RADXS selection on stars both within and outside the defined WD region, enclosed by two green lines (see text for details). Lastly, in panel (s2), a direct comparison between real and artificial stars reveals no distinct decline in the number of artificial stars below $m_{\rm F606W}\sim30.1$.  }
\label{fig:sel}
\end{figure*}

Following the approach by  \citet{2019MNRAS.488.3857B,2023MNRAS.518.3722B}, in Fig.\,\ref{fig:sel}, we show the impact of our progressive selection criteria on ASs, and then apply the same criteria to real sources. In each panel, the top-right corner is labelled either with an (a) for ASs or with an (r) for real sources. The aim of these selections is to strike a balance that maximizes the inclusion of well-measured WD members of $\omega$\,Cen while minimizing the inclusion of spurious or poorly measured detections. It is important to note that in the subsequent analysis, we exclusively utilize the long exposures, as they provide the necessary depth to study the WD CS.

The light blue and dark blue shaded areas represent the 5$\sigma$ and 3$\sigma$ regions, respectively, where $\sigma$ denotes the background noise as measured by  \citet{2019MNRAS.488.3857B} in both filters. For our analysis, we only consider sources above the 3$\sigma$ limits as significantly detected.

Panel\,(a1) displays all artificial sources as inserted (in magenta) and as recovered (blue dots). In panel\,(a2), sources are limited to those within areas observed in at least $\sim$40\% of the F814W and F606W images, resulting in a significant reduction of the field of view used for the investigation.

Panel\,(a3) further restricts sources to regions where the rmsSKY matches the noise within empty sky patches, indicating areas suitable for detecting faint objects. Panel\,(a4) excludes all ASs recovered outside the region enclosed within the two green lines shown in panel\,(a3).

In panel\,(c1), instead of a CMD, the magnitude versus completeness curve is presented (black line). We also display the "good" completeness $c_g$ in blue, which represents the completeness calculated within the "good" regions based on the rmsSKY value \citep[see][for details]{2008ApJ...678.1279B}. This panel indicates that sources passing these selections are 50\% complete down to $m_{\rm F606W}$=29.45 and 25\% complete at $m_{\rm F606W}$=30.04.

Panel\,(a5) displays the result after applying the RADXS parameter to reject non-stellar objects. Panel\,(c2) shows the completeness curves after the final RADXS selection.

The same selections applied to ASs are then applied to observed real sources in the top panels of Fig.\,\ref{fig:sel}. Panel\,(r5) shows the final sample for $\omega$\,Cen’s WD candidates.

Panel\,(s1) displays all stars that meet the selection criteria, including those outside the area defined by the two thin green lines in panel\,(a3). The stars outside this region will be utilized to model field contamination within the WD region in the CMD. The fiducial line defined in Section\,\ref{Section4} (and shown in magenta) well represents the mean observed CMD location for WD CS of $\omega$\,Cen.

Panel\,(s2) shows real stars and ASs that meet all selection criteria: ASs --as selected in panel\,(a5)-- are shown in orange, while real sources --as selected in panel\,(r5)-- are shown in blue. This comparison suggests that the majority of the observed real WD CS might not extend to magnitudes as faint as those of the recovered ASs. This could indicate that the peak of the WD CS LF of $\omega$\,Cen has potentially been reached and surpassed.

\section{Proper motion decontamination}\label{Section6}
We also investigated whether the PMs can help in disentangling field objects found within the WD region of $\omega$\,Cen's CMD and cluster members. To do this, we evaluated PMs following the approach by \citet{2023MNRAS.518.3722B} i.e. combining the bulk of the first three epochs ($\sim$2016.1) to obtain averaged positions for the sources, which are then compared to their averaged positions as measured in the last three epochs of the data ($\sim$2018.1). In the following analysis, we will consider only sources shown in green in panel\,(s1) of Fig.\,\ref{fig:sel} but for which it was possible to estimate the PM. 

\begin{figure*}
\centering
\includegraphics[width=\textwidth]{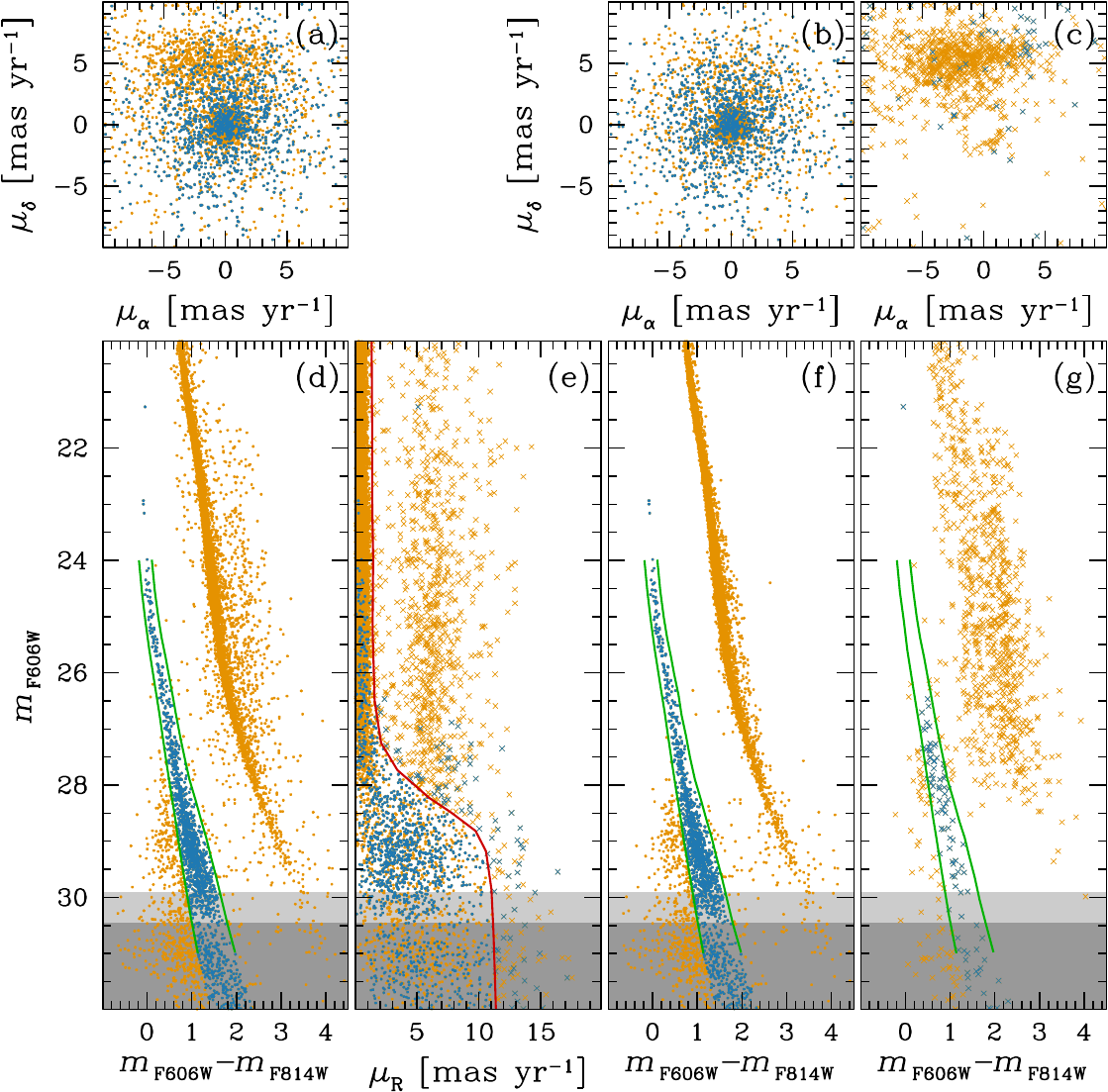}
\caption{(a)-(b)-(c) Vector point diagrams for the samples shown in the corresponding panels below. (d) CMD of the sources in panel (s1) of Fig.\,\ref{fig:sel}. In all panels, sources within the WD region (between the two green lines) are represented in blue, while all other sources are shown in orange. The light-grey and dark-grey shaded area indicate the $5\sigma$ and $3\sigma$ limit, respectively, of significant detection for the sources of interest. (e) One-dimensional PM, $\mu_{\rm R}$, as a function of the $m_{\rm F606W}$ magnitude. Bright stars (down to $m_{\rm F606W} \sim 28$) exhibit a $\mu_{\rm R}$ distribution with a tight dispersion ($<1.5$ mas yr$^{-1}$) along with a tail displaying a much broader dispersion. We arbitrarily define two regions, indicated by the red line: one enclosing the bulk of the $\mu_{\rm R}$ values at different magnitudes (indicated by filled circles), and the other containing objects with larger $\mu_{\rm R}$ (indicated by crosses). (f)-(g) CMDs for the stars within and beyond the red line defined in panel\,(e). Neither of the two CMDs solely consists of members or field objects (see text for details).}
\label{fig:pm}
\end{figure*}

The PM analysis is presented in Fig.\,\ref{fig:pm}. We colour-code in blue stars surviving the WD selection defined by the two thin green lines, and in orange all other stars.

In panels (a), (b) and (c) we present the vector-point diagrams (VPDs) of the sources in our sample. Since PMs are calculated relative to the cluster's overall motion, the cluster members' distribution in the VPD is centred at (0,0). Panels (d), (f) and (g) display CMDs. Panel (e) from the left shows the observed one-dimensional PM ($\mu_{\rm R}$, obtained by summing in quadrature the PM in the two directions) plotted against $m_{\rm F606W}$. Among brighter stars, we observe a narrow distribution in $\mu_{\rm R}$ for $\omega$\,Cen's members, predominantly clustered well below $\mu_{\rm R}<1.5$ mas yr$^{-1}$, while a broader tail extends towards higher $\mu_{\rm R}$, peaking between $2<\mu_{\rm R}<10$ mas yr$^{-1}$, indicative of field objects. However, as we approach a magnitude $m_{\rm F606W} \sim 28$, the random errors in positional measurements, compounded over the first and last three epochs, become significant for fainter stars. By $m_{\rm F606W}\sim28.5$, it becomes notably challenging to disentangle cluster members and field objects. We defined a PM selection consistent with the PM errors at different magnitudes represented by a red line. Stars satisfying this selection are shown as filled circles, while stars not passing this selection are indicated with crosses. The subsequent panels display the VPD and CMD for sources positioned to the left or right of the red-line criterion, maintaining the blue colour code for WD candidates. The PM selection effectively separates outliers and objects with large PMs. However, it struggles to distinguish between WDs and field objects fainter than $m_{\rm F606W}\sim28.5$. This challenge arises mainly due to the minimal separation between field and cluster members, which is much smaller than measurement errors at magnitudes fainter than $m_{\rm F606W}\sim28.5$.

Since PMs are not effective in distinguishing between WDs and field objects for faint sources, we decided not to use PMs in our analysis.

\section{The corrected WD CS LF}\label{Section7}

\begin{figure*}
\centering
\includegraphics[width=\textwidth]{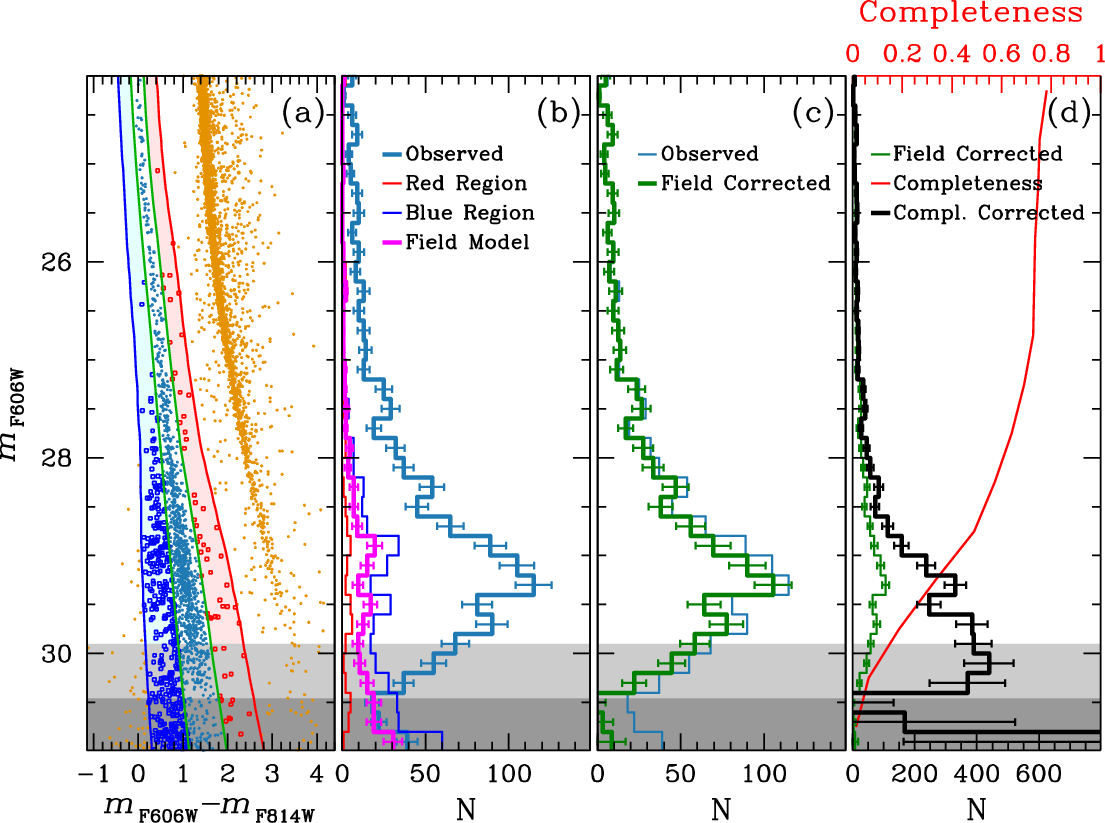}
\caption{(a) The CMD is partitioned into three regions along the WD CS. The azure-shaded region indicates sources that pass the selection on the blue side of the WDs, while the red-shaded region represents those on the redder side. (b) WD CS LF, displaying histograms depicting the number of sources per magnitude interval for observed stars within the WD region and for stars in the two shaded regions. The magenta histogram represents our model for field distribution. (c) The WD CS LF derived by subtracting the field model from the observed WD CS LF. (d) The observed field-corrected WD CS LF, adjusted for completeness, is depicted in black. 
Grey-shaded areas indicate the 5- and 3-$\sigma$ levels,  where below 3-$\sigma$ findings and completeness become unreliable. Errors were linearly propagated and then corrected for completeness.}
\label{fig:lf}
\end{figure*}

We applied a correction to the observed WD CS LF to mitigate the effects of field contamination by unresolved blue galaxies, following the method outlined by \citet{2023MNRAS.518.3722B}. The process is presented in Fig.\,\ref{fig:lf} and described as follows. In panel\,(a), we show the CMD of sources identified in panel\,(s1) of Fig.\,\ref{fig:sel}. Using the two lines defined in Fig.\,\ref{fig:sel}, we establish on the CMD what we term the \textit{WD-region}. Additionally, we define two other regions with the same colour width at each magnitude of the WD-region, one at bluer colours and the other at redder colours, referred to as \textit{Blue-} and \textit{Red-regions}, respectively. These regions are symmetrically offset from the WD-region by a fixed colour interval at each magnitude.

We then counted the sources within each of these three regions and generated in panel\,(b) the LFs for each one of them, with error bars representing statistical Poisson errors. The number of contaminants within the CMD WD region is determined as the average of the number of objects observed in the \textit{Blue-} and \textit{Red-regions} at various magnitudes. This model is shown in magenta, with corresponding errors estimated through linear propagation of Poisson noise.

In panel\,(c), we compare the observed LF with the resulting WD LF corrected for the field-contamination model. The field-corrected LF is obtained by subtracting the field model from the observed LF, with errors propagated linearly.

Finally, in panel\,(d), we present the completeness-corrected and field-corrected WD CS LF for $\omega$\,Cen. The errors on this LF were corrected for completeness using a simple approximation involving linear propagation of the errors.

The WD CS LF exhibits a peak at an estimated magnitude of $m_{\rm F606W} = 30.1\pm0.2$, followed by a rapid decline leading to zero. Despite the low completeness at these faint magnitudes, which falls well below the commonly accepted limit of 50\%, we can consider this result valid as the peak exceeds the 3$\sigma$ threshold (dark grey area) that we have set as the lower limit for valid measurements.

\section{Theoretical interpretation of the WD LF}\label{Section8}
As briefly summarized in the introduction, $\omega$\,Cen hosts a very complex population of  stars, composed of several subpopulations that show themselves as multiple sequences in CMDs displayed in appropriately chosen filter combinations. These multiple sequences are originated by the range of initial chemical composition -- and possibly age-- of the cluster's stars.   

Spectroscopic studies have determined a range of iron abundances $\sim -2.2 <$[Fe/H]$< \sim -$0.6 \citep[see, e.g.][]{jp10, marino11,2024ApJ...970..152N}, with the main component characterized by [Fe/H]$\sim -$1.7, and the canonical globular cluster light-element abundance anticorrelations (e.g. the O-Na anticorrelation) present at all [Fe/H]. In addition, there is a sizable component of stars with [Fe/H]$\sim-$1.3 and an enhanced helium mass fraction, i.e. $Y\sim$0.36-0.40 \citep[see, e.g.,][]{p05, 2012AJ....144....5K}. From the point of view of the stellar ages, the situation is much less clear. For example, \citet{s05b} found an age spread $\Delta$t $\leq$2~Gyr from their analysis of the CMD of the cluster subgiant branch (SGB) region, a result in agreement with the analysis by \citet{calamidaage} based on deep near-infrared photometry,
and \citet{tailo} derived a negligible $\Delta$t from the joint analysis of the HB and SGB.
On the other hand, the studies by \citet{v07} and \citet{villanovaage} of the SGB provided 
$\Delta$t equal to several Gyr.

\begin{figure}
\centering
\includegraphics[width=\columnwidth]
{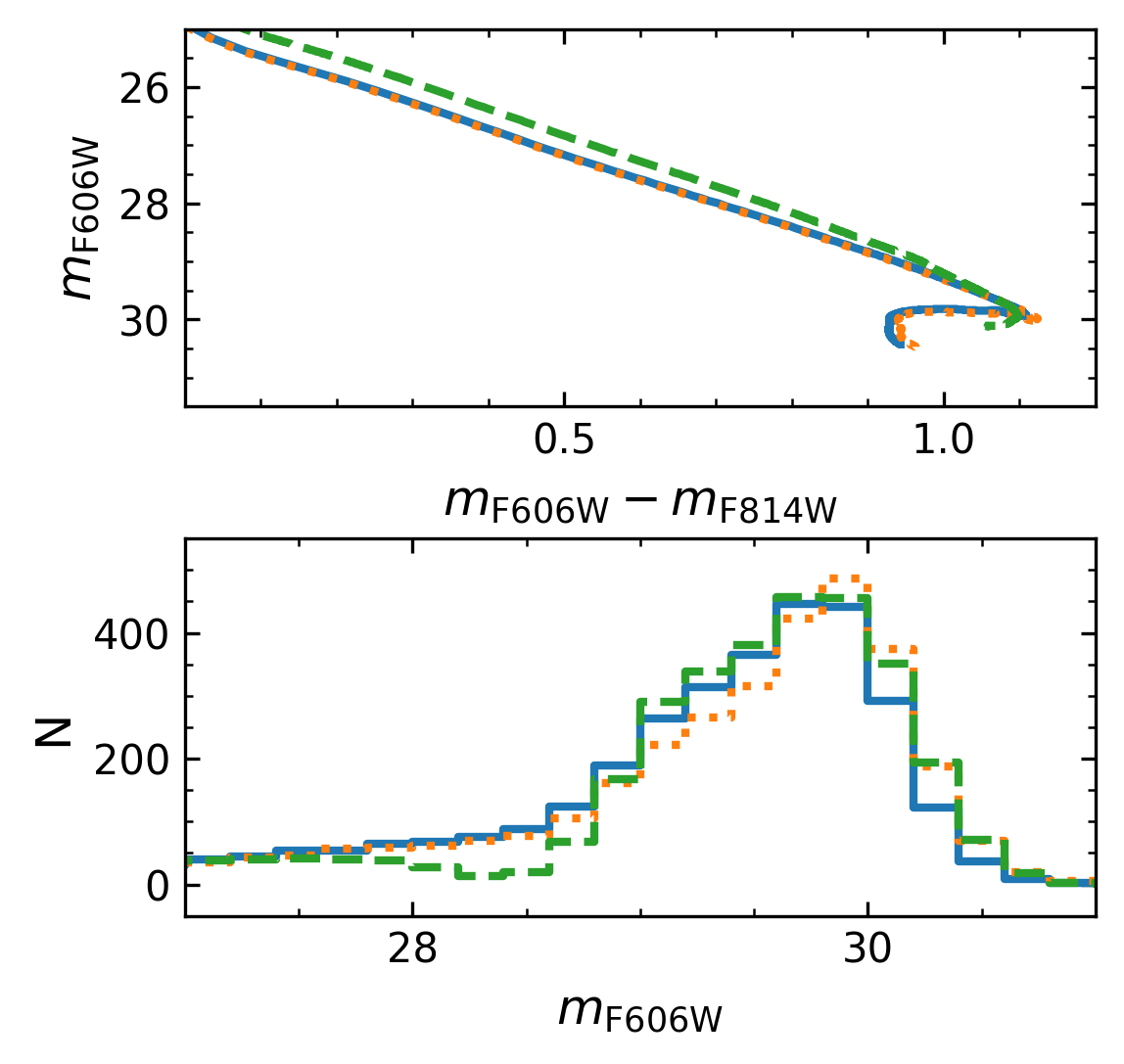}
\caption{{\sl Upper panel:} Two 12~Gyr WD isochrones for the progeny of He-normal cluster subpopulations with [Fe/H]=$-$1.9 and $-$0.9 (blue solid and orange dotted lines, respectively), and a 12~Gyr WD isochrone for the progeny of the $Y$=0.4 cluster subpopulation (green dashed line). See text for details. {\sl Lower panel:} LFs calculated from the three isochrones in the upper panel (same line styles and colours). The number of stars in each LF is the same, approximately equal to the total number of 
objects in the cluster LF.}
\label{fig:lfcmdtcomp}
\end{figure}

Here we have investigated theoretically the observed WD CS and the LF of Fig.~\ref{fig:lf} in light of this complex chemical abundance (and possibly age) distribution, by employing sets of WD cooling tracks and isochrones.
To model the CO-core WDs from progenitors not belonging to the very He-enhanced $Y$$\sim$$0.4$ 
cluster subpopulation we have employed the 
BaSTI-IAC WD tracks by \citet{bastiiacwd} with hydrogen envelopes
and metal-poor progenitors, calculated with the 
\citet{opadeg} electron conduction opacities, complemented by 
the \citet{ifmr} initial-final mass relation and progenitor lifetimes from the $\alpha$-enhanced \citet{bastiiacaen} models to calculate the corresponding WD isochrones.
For the WDs produced by the $Y$$\sim$0.4 subpopulation, given the lack of He-core WD tracks and progenitor models with $Y\sim$0.40 in the BaSTI-IAC database, we employed the models by \citet{wdhe} for $Y$=0.40 and a metal mass fraction $Z$=0.001, close to the value appropriate for the He-rich stars in the cluster (characterized by [Fe/H]$\simeq-$1.3). \citet{wdhe} calculations follow the evolution from the main sequence to the WD stage of models with masses from 0.6$M_\odot$ to 2 $M_{\odot}$ and $Y$=0.40, and provide therefore also a theoretical initial-final mass relation and progenitor lifetimes for these WDs, enabling us to calculate WD isochrones for the $Y\sim$0.4 cluster subpopulation. According to these calculations, MS stars with 
initial masses up to 0.65$M_{\odot}$ produce He-core WDs, while more massive objects produce CO-core WDs.
At ages around 10-13\,Gyrs WD isochrones calculated from these models predict along the bright CS He-core WDs with masses equal to 0.44-0.46$M_{\odot}$, consistent with the results by 
\citet{2013ApJ...769L..32B}.
A couple of points must be noted. The first one is that 
\citet{wdhe} calculations (both progenitors and WDs) do not employ the same physics inputs of the BaSTI-IAC, although some of the main inputs are in common (for example the equation of state for the model WD cores). The second point is that \citet{wdhe} calculate progenitor models only up to 
2$M_{\odot}$, that produce a CO-core WD model with mass equal 
to 0.81$M_{\odot}$. More massive WDs are therefore missing 
from the isochrones calculated with \citet{wdhe} models.

Figure~\ref{fig:lfcmdtcomp} compares three 12~Gyr isochrones and the corresponding LFs (the result of this comparison does not depend on the chosen age, at least for ages older than a few Gyr), after applying the distance modulus and extinctions assumed for $\omega$\,Cen. Throughout our analysis we use 
$(m-M)_0$=13.67 determined by \citet{bv} as an average of $Gaia$ Early Data Release~3 parallax  
and kinematic distances, $HST$ kinematic distance, and 26 literature estimates (listed in their Table~B.1) based on pulsating variables, eclipsing binaries, CMD fitting, and 
tip of the red giant branch. The adopted value 
agrees well with the eclipsing binary distance by 
\citet{thomp}.

For the reddening we employed 
$E(B-V)$=0.12 \citep{2012AJ....144....5K},  
close to the average value $E(B-V)$=0.11 from multiple sources recommended by \citet{lub}, and consistent with the mean value determined by \citet{calamidaredd} using Str\"omgren photometry. 
Using this value of $E(B-V)$ we have applied extinction
corrections to the F606W and F814W filters dependent on the model $T_{\rm eff}$ calculated as in \citet{2005MNRAS.357.1038B}.

These LFs (with the same bin size of the 
observed cluster LF) and all other theoretical LFs discussed later have been calculated 
using a power-law mass function for the progenitors with a Salpeter exponent $x=-2.3$ and include with a Monte Carlo technique the photometric error law 
for the $\omega$\,Cen CS derived from the AS analysis, as described in \citet{2023MNRAS.518.3722B}.
Two isochrones have been calculated for CO-core WDs from He-normal progenitors with [Fe/H]=$-1.90$ and $-0.90$, respectively, while the third isochrone has been calculated for the progeny of the $Y\sim$0.40 population. This isochrone includes He-core models down to $m_{\rm F606W}\sim$28.5, and CO-core models at fainter magnitudes.

The isochrones for the two He-normal populations almost completely overlap, 
and are very similar at their faint end, the age 
indicator of the population. This is a consequence of the fact that progenitors' metallicity variations affect the WD cooling speed mainly due to the variation of the $^{22}$Ne abundance in the core that affects the energy released by $^{22}$Ne diffusion and distillation \citep[see, e.g.][and references therein]{db, altdiff, blouin, bastiiacwd, dist}; however, at the low metallicities 
of globular clusters, these effects are negligible because of the low abundance of $^{22}$Ne in the WD cores. 
Moreover, the faint end of the isochrone, where the colours turn to the blue, is populated by the more massive WDs (with 
decreasing radii), coming from progenitors whose lifetime is much shorter than the isochrone age. As a consequence -- remembering that at each point along a WD isochrone 
the sum of the cooling age of the WD evolving at that point and its progenitor lifetime is constant and equal to the isochrone age-- the corresponding magnitudes are set by the cooling times of the WD progeny, and are unaffected by small changes in the progenitor lifetimes due to variations in their metallicity.
The \lq{hook\rq} at the blue end 
of the isochrones is caused by the 
most massive objects (masses $\sim 1-1.1 M_{\odot}$ that have cooled down faster at 
this isochrone age, due to an earlier onset of crystallization and phase separation, and are 
slightly fainter compared to the less massive WDs.

The isochrone for the He-rich population is redder for most 
of its magnitude extension, because it is populated by lower mass objects, 
and eventually approaches the other two isochrones 
when the evolving masses become similar. The magnitude of the bottom end of this isochrone is slightly fainter, and the colour extension shorter, this latter effect is due to the lack of models 
with mass above 0.81$M_{\odot}$ (the other two isochrones include WD models up to 1.1$M_{\odot}$).

Despite these differences, once photometric errors are included, the LFs calculated from the three isochrones are almost equivalent, especially when considering the errors in the observed star counts of the observed LF. 

As mentioned in the Introduction, \citet{2013ApJ...769L..32B} near-UV and B $HST$ photometry of the bright part of the cluster CS has demonstrated that there are two parallel sequences. According to their analysis  --which also made use of \citet{omegahb} study of the extreme blue part of the cluster HB-- the redder sequence is populated essentially by He-core WDs with mass around 0.45$M_{\odot}$, the progeny of the He-rich sub-population, while the blue sequence hosts the standard CO-core WDs with mass around 0.55$M_{\odot}$ expected to populate the 
bright CS of globular clusters.

The bright part of our optical CS however does not show a split sequence, due to the small sample of objects and especially the low sensitivity of the (F606W$-$F814W) colour to $T_{\rm eff}$ at high temperatures, as shown by the following test.  

\begin{figure}
\centering
\includegraphics[width=\columnwidth]
{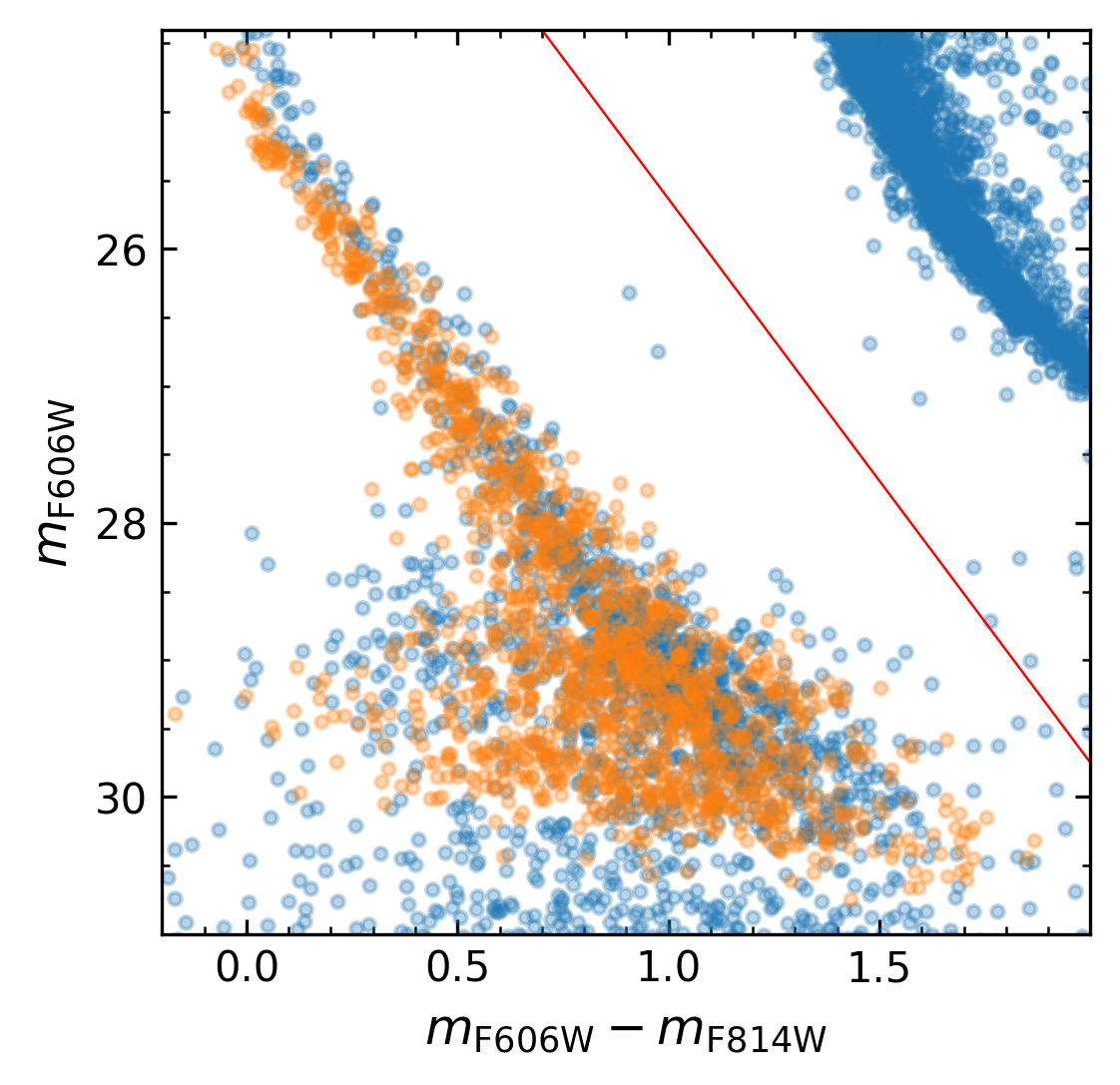}
\caption{Cluster CS (blue filled circles) compared to a 12~Gyr synthetic CMD of the cluster WD population (orange filled circles). The red line is used to define the sample of stars used in our analysis. See text for details.} 
\label{fig:syntcomp}
\end{figure}

Figure~\ref{fig:syntcomp} displays a qualitative comparison  between the cluster cooling sequence and a 12~Gyr (a representative age) synthetic CS  made of two components (we used for the synthetic CS the cluster distance modulus and extinction discussed before). The main component, which includes 89\%  of the synthetic population, represents the WDs produced by all cluster sub-populations other than the $Y\sim$0.4 one. Given the result of the isochrone comparison in Fig.~\ref{fig:lfcmdtcomp}, we employed a single WD isochrone  for a representative [Fe/H]=$-$1.7 to produce this WD population. The remaining 11\%  of the synthetic CS has been calculated using the isochrone calculated with \citet{wdhe} $Y$=0.4 progenitors and WDs.  These percentages come from the analysis of  the radial distribution of the cluster sub-populations presented in \citetalias{2024A&A...688A.180S}, taking into account that the observed field is  located at a radial distance from the centre equal to 2.6 times the half-light radius. 

The magnitudes along the synthetic CS (which include the photometric errors as determined from the artificial star analysis) have been calculated as described in  \citet{2023MNRAS.518.3722B}. Here we also account for the effect of completeness (we display only objects that pass the completeness test), considering how the completeness fraction varies with $m_{\rm F606W}$ according to the AS analysis. 

The number of synthetic objects is 1498, the same as on the observed CS, defined in this comparison as the sequence of cluster's stars to the left of the straight line in the figure, with colours larger than $-$0.5~mag and F606W magnitudes lower than 30.5.

The synthetic sequence appears in good overall qualitative agreement with the observations, in terms of shape and colour spread at a given magnitude, apart from the brighter magnitudes below $m_{\rm F606W}\sim$26, where the synthetic CS appears to be slightly bluer than the observations.  The synthetic population confirms that with this observed CMD we do not expect to see a well-defined bimodal CS even at the brightest magnitudes where the photometric error is small.

\begin{figure}
\centering
\includegraphics[width=\columnwidth]
{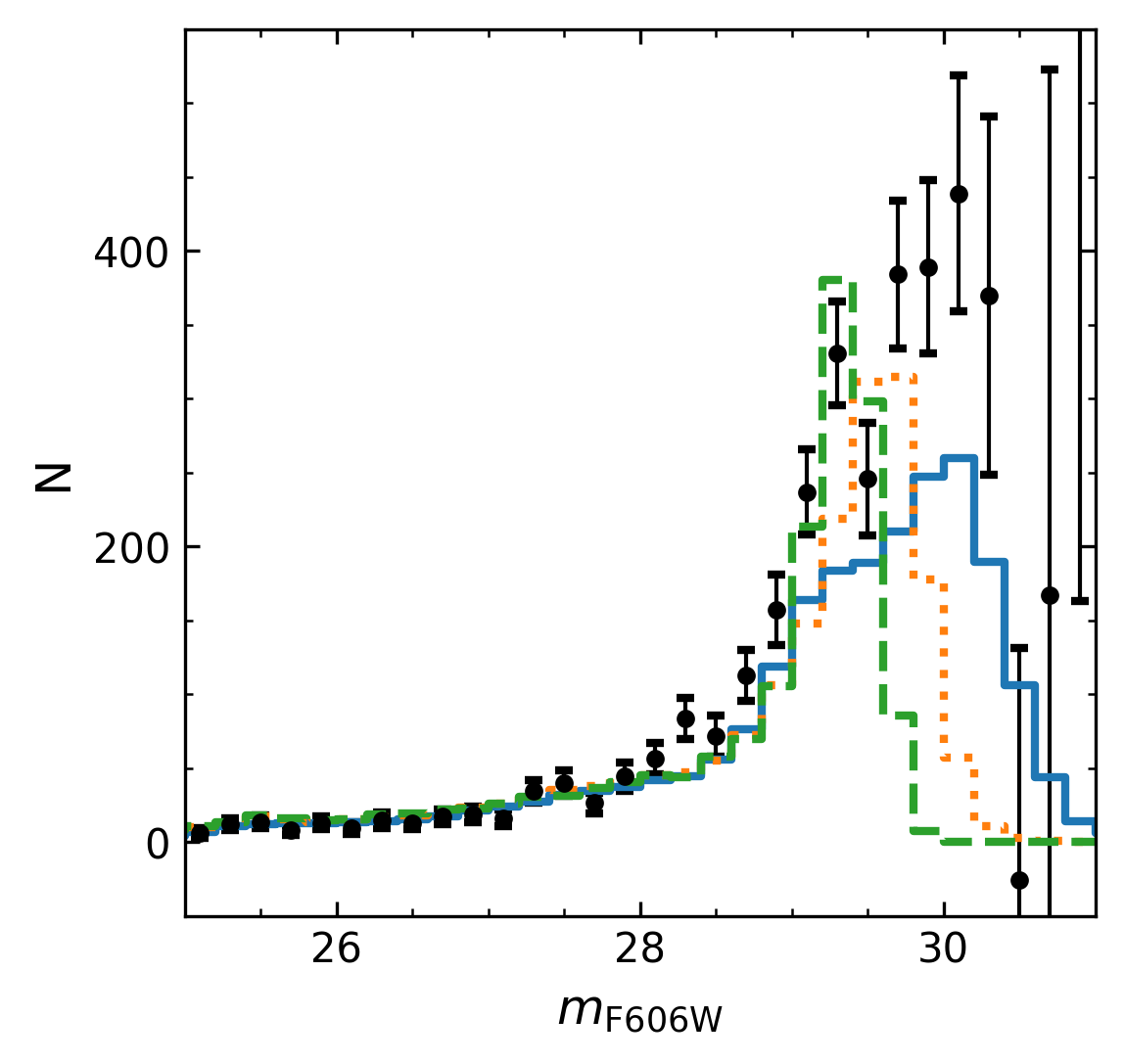}
\caption{Cluster LF (filled circles with error bars) 
compared to theoretical LFs for ages equal to 13 (blue solid line), 10 (orange dotted line) and 8~Gyr (green dashed line). See text for details.}
\label{fig:agerange}
\end{figure}

Before trying to interpret the cluster LF in terms of age and age spread of the various subpopulations, we compare it in 
Fig.~\ref{fig:agerange} with three theoretical LFs 
for 8, 10 and 13~Gyr respectively, calculated for He-normal populations and [Fe/H]=$-$1.71. The theoretical LFs have been normalized to have the same 
total number of objects as in the empirical LF for 
$m_{\rm F606W} <$ 28.0.
As shown by Fig.~\ref{fig:lfcmdtcomp}, a LF calculated using a single initial chemical composition for the progenitors is appropriate to represent 
the whole range of chemical compositions of the cluster subpopulations (including the $Y\sim$0.40 subpopulation). 

The width of the peak at the faint end of the LF encompasses the theoretical LFs for ages between 8 and 13 Gyr, but neither a single age nor a combination of ages between 8 and 13 Gyr can match the observed star counts between $m_{\rm F606W}$ of 29.5 and 30.3. This is very likely an indication that 
the mass distribution of the WDs along the cooling sequence predicted 
by our choice of a Salpeter progenitor mass function is inappropriate.
\citep[see, e.g., the discussion in][]{2023MNRAS.518.3722B}, due to the 
cluster dynamical evolution. A more top-heavy mass function would increase the relative number of massive WDs and therefore 
increase the number counts across the region of the peak of the LF.

\begin{figure}
\centering
\includegraphics[width=\columnwidth]
{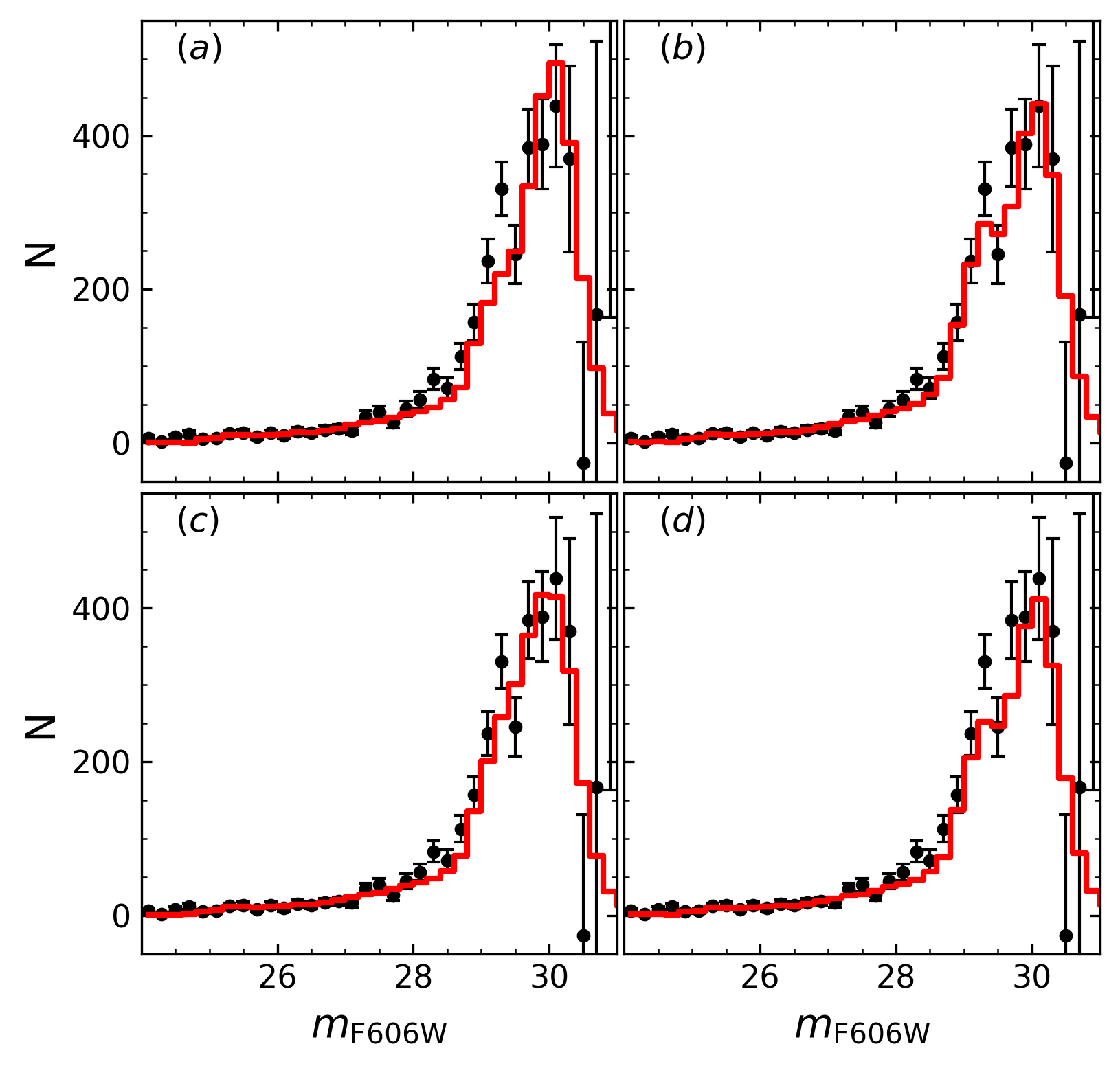}
\caption{Cluster LF compared to theoretical LFs calculated with an exponent $x=-1.6$ for the WD progenitor mass function, and various age combinations. {\sl Panel a}: single age 13~Gyr old population; {\sl Panel b}: 12\% of the population 8~Gyr old, 88\% is 13`Gyr old; {\sl Panel c)}:  4\% of the population is 8~Gyr, 16\% is 10~Gyr old, and 80\% is 13 Gyr old; {\sl Panel d)}: 8\% of the population is 8~Gyr old, 8\% is 12~Gyr old, and 84\% is 13~Gyr old. 
}
\label{fig:sfh}
\end{figure}

Figure~\ref{fig:sfh} indeed shows how an exponent $x=-$1.6 
for the progenitor mass function produces theoretical LFs (normalized as described before) that better match the star counts in the region of the peak of the LF, for various selected combinations of ages.
The maximum age range compatible with the observations is 5~Gyr, 
with 8 Gyr as the minimum age to match the rising branch of the peak region of the LF, while an age of 13 Gyr is required to match the faint, descending branch of the peak region.
A single-age population for the cluster provides a good fit to the data, also taking into account that the local maximum of the LF 
centred at $m_{\rm F606W}=29.3$ (not matched by the theoretical LF) is very likely a statistical fluctuation; when considering the error bars, the star counts in this magnitude bin are different from the values at the two adjacent bins by less than 2$\sigma$.

That said, an age range among the cluster population cannot be ruled out, so long as the large majority of stars (at least $\sim$80\%) in the cluster have an age of 13~Gyr, as shown by the three selected examples in Fig.~\ref{fig:sfh}. A high fraction of stars with an age of 13~Gyr is required to match the two points at $m_{\rm F606W}$=30.1 and 30.3, while the fraction of stars with ages as low as 8~Gyr has to be small, less than $\sim$10\%, otherwise the ratio between the star counts around $m_{\rm F606W}=29$ and the number of stars around $m_{\rm F606W}=30$ becomes too high compared to the observed LF.  The quality of the matches between theory and observations in Fig.~\ref{fig:sfh} is very similar for all four representative cases, and {\sl ad-hoc} tweaks of the mass function can improve the match in each of them.  Therefore an age range between 8 and 13~Gyr among $\omega$\,Cen stars cannot be ruled out from the analysis of the WD CS, but if stars younger than 13\,Gyr do exist, they should not make up more than $\sim$20\% of the total, with less than 10\% can having an age as young as 8~Gyr.

\section{Summary}\label{Section9}
We have presented our study of the complete WD CS in $\omega$\,Cen, the primary objective of the HST GO-14118+14662 program. We have produced the CMD and corresponding completeness-corrected LF of the WD\,CS, which displays a peak at the termination of the CS, located at a magnitude $m_{F606W}=30.1\pm0.2$.

We have created a synthetic WD CS for $\omega$\,Cen, consisting of a main component made of CO-core objects progeny of the He-normal cluster subpopulations (89\% of the WDs) and a component made of CO-core and He-core objects produced by the He-rich cluster subpopulation. This synthetic CS aligns well with observational data, except at brighter magnitudes where the synthetic sequence appears slightly bluer.

Our analysis has shown that the chemical complexity of the cluster stellar population has a minor impact on the theoretical interpretation of the observed LF. For a fixed age, LFs calculated for CO-core WDs with varying progenitors' initial chemical compositions, and for He-core WDs produced by the helium-rich cluster subpopulation are very similar, particularly when considering the observational errors. 

We found that a single-age population can match overall the observed LF, but an age range cannot be entirely ruled out using just the WD LF, given the uncertainties in the present-day WD mass function. LF comparisons suggest that $\omega$\,Cen's star formation history could potentially span an age range up to $\sim$5 Gyr,  however, the majority of stars (at least 80\%) must be approximately 13 Gyr old, and only a small fraction (less than 10\%) could potentially have ages be as young as 8 Gyr.

Further studies, analysing simultaneously the WD LF together with both spectroscopic and photometric data of the previous evolutionary phases, are essential to fully understand the formation and evolution of this extreme globular cluster. In particular, we note that observations of these very same fields are approved and scheduled with JWST under program \href{https://www.stsci.edu/jwst/science-execution/program-information?id=5110}{GO-5110} (PI: Bedin). 

\begin{acknowledgements}
Michele Scalco and Luigi Rolly Bedin acknowledge support by MIUR under the PRIN-2017 programme \#2017Z2HSMF, and by INAF under the PRIN-2019 programme \#10-Bedin.
\end{acknowledgements}

\bibliographystyle{aa}
\bibliography{main.bib}

\end{document}